# Ideal Databases

Vadim Tropashko
Vadim.Tropashko@or..cle.com
vadimtro@gm..l.com

___________________________________________________________________

From algebraic geometry perspective database relations are succinctly defined as Finite Varieties. After establishing basic framework, we give analytic proof of Heath's theorem from Database Dependency theory. Next, we leverage Algebra-Geometry dictionary and focus on algebraic counterparts of finite varieties – [polynomial] Ideals. It is well known that intersection and sum of ideals are lattice operations. We generalize this fact to ideals from different rings, therefore establishing that algebra of ideals is Relational Lattice. The final stop is casting the framework into Linear Algebra and traverse to Quantum Theory.

___________________________________________________________________

INTRODUCTION

Constraint databases were promoted since the 1990s by Paris Kanellakis with conventional wisdom of representing relations as semi-algebraic sets. This insight was motivated by Tarski-Seidenberg theorem, which asserts that projection of semi-algebraic set is semi-algebraic. Together with folklore knowledge that the class of all semi-algebraic subsets is closed under finite unions and intersections, taking complement, inverse image by a polynomial mapping, and Cartesian product it becomes obvious that semi-algebraic sets fit nicely into Codd's relational algebra.

This article shifts the emphasis from inequalities to equalities with mathematical inspiration from Algebraic Geometry. The basic geometric object is *affine variety* — a system of polynomial equations. Affine variety is a set of tuples in $R^n$ while our focus on database applications prompts that we must narrow our scope to finite varieties, when polynomial system has finite number of roots.

Unlike semi-algebraic sets, polynomial varieties are not closed under projection. One needs to treat projection of algebraic set as geometric projection combined with elimination of variables via Zariski closure. Such combined operation transforms algebraic set into algebraic set. Then, canonical algebra-geometry dictionary from classic textbook [1] (page 214) provides a recipe for algebra of finite varieties resembling Codd's relational algebra. This development takes up the first part of the article and is concluded with analytical proof of Heath's theorem.

In the second part we we shift the focus from varieties to dual algebraic objects — [polynomial] *ideals*. Unlike varieties ideals describe not only the set of roots where polynomial system vanishes, but root multiplicities as well. In database terminology ideals are multi-relations. We make quick comparison of naive database model of multi-relation with ideal and explain why ideals capture enough information to give rise to a consistent algebra. We conclude this section with main result generalizing well-known fact that intersection and sum of ideals are lattice operations. We illustrate those findings with cartoon version of RDBMS implemented in CoCoA computer algebra system.

In the final section we study polynomial ring with Linear Algebra methods borrowed from [2]. This is quite refreshing perspective: attributes of a database relation can be viewed as commuting linear operators. Attribute values are eigenvalues of corresponding operator. The entire picture gets distinct quantum mechanical flavor, where database attribute is essentially an observable.

1. FINITE VARIETIES

Basic object of algebraic geometry is *affine variety* — a system of multivariate polynomial equations. Affine variety is a set of points in $R^n$ (or $C^n$ )[1] and our database focus prompts that we must narrow our scope to finite varieties, when polynomial system has a finite number of roots.

Our first task is to describe how to construct a variety out of any given database relation. Database relations are assembled from smaller pieces by joining attribute values into a tuple, then unioning the tuples. This prompts the need in the two fundamental operations over varieties.

1. Set intersection of two varieties $V$ and $W$ is a set which is, again, a variety. The defining set of polynomials for $V \cap W$ is a union of the polynomial constraints systems defining the $V$ and $W$.

2. Set union of two varieties $V$ and $W$ is a set which is, again, a variety. The defining set of polynomials for $V \cup W$ is a set of all pairwise products of the polynomials from the systems defining $V$ and $W$.

Now, we can exhibit a variety corresponding to any relation. Consider an unary relation with a single tuple

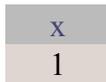

Geometrically, it corresponds to a single point on the $x$ axis, so it becomes immediately obvious what equation defines it:

$$x - 1 = 0$$

---

[1] Certainly, we have to work with nice domains such as algebraically closed field of complex numbers $C$, or real numbers $R$, at least.

Likewise, the relation

| y |
|---|
| 1 |

is defined by

$$y - 1 = 0$$

Next, we construct join of these two relations – a binary relation with one tuple

| x | y |
|---|---|
| 1 | 1 |

Join is set intersection, and intersection of varieties gives us the system of defining equations

$$x - 1 = 0$$
$$y - 1 = 0$$

Let's expand our example and add one more attribute to the relation:

| x | y | z |
|---|---|---|
| 1 | 1 | 1 |

Our system of equations grows with one more constraint:

$$\begin{aligned} x - 1 &= 0 \\ y - 1 &= 0 \\ z - 1 &= 0 \end{aligned} \quad (2.1)$$

Now, knowing how to construct "single tuple varieties", we are ready to move onto relations with more than one tuple. This is accomplished via union. Consider a ternary relation:

| x | y | z |
|---|---|---|
| 1 | 1 | 1 |
| 2 | 1 | 2 |

which is a union of already familiar relation

with

| x | y | z |
|---|---|---|
| 1 | 1 | 1 |

| x | y | z |
|---|---|---|
| 2 | 1 | 1 |

To build the union of varieties we need a polynomial system defining the second variety

$$x-2=0$$
$$y-1=0 \quad (2.2)$$
$$z-1=0$$

Taking set of all pairwise products of all the polynomials from the systems 2.1 and 2.2 above we obtain the following polynomial system for the union:

$$(x-1)(x-2)=0$$
$$(y-1)(x-2)=0$$
$$(z-1)(x-2)=0$$
$$(x-1)(y-1)=0$$
$$(y-1)(y-1)=0$$
$$(z-1)(y-1)=0$$
$$(x-1)(z-1)=0$$
$$(y-1)(z-1)=0$$
$$(z-1)(z-1)=0$$

At this stage the complexity of this polynomial system seems to be discouraging, but we have grossly over specified the system of constraints, because not all of these equations are independent. Groebner basis is ubiquitous method to find a set of independent polynomials. Executing `GroebnerBasis` command in a typical Computer Algebra system would output much smaller set of equations:

$$x^2-3x+2=0$$
$$y-1=0$$
$$z-1=0$$

As an afterthought, this result is obvious. The first equation constraints $x$ to being either $1$ or $2$, the second equation asserts that $y$ is equal to $1$, while the third one asserts $z=1$. It is satisfying to know the general method, though.

With this technique (applying the union rule and, consequently, reducing the system with Grobner basis) we can proceed and find a variety corresponding to the relation with 4 tuples:

| x | y | z |
|---|---|---|
| 1 | 1 | 1 |
| 2 | 1 | 1 |
| 3 | 2 | 1 |
| 3 | 2 | 2 |

$$-6+11x-6x^2+x^3=0$$
$$-4+3x-x^2+2y=0$$
$$3-x-3z+xz=0$$
$$2-3z+z^2=0$$

This is, again, the exact output of GroebnerBasis command, but for human reader factorizing these polynomials makes them more intuitive

$$(x-3)(x-2)(x-1)=0$$
$$-4+3x-x^2+2y=0$$
$$(x-3)(z-1)=0$$
$$(z-2)(z-1)=0$$

The first and the last equations are univariate. The first equation asserts that values $x$ are restricted to the set $\{1,2,3\}$. Likewise, the last constraint claims $z \in \{1,2\}$. The third equation asserts that either $x=3$ or $z=1$. What is the second constraint?

Since there is only one monomial containing $y$ (and it is limited to the first power), the equation can be rewritten as

$$y=(-4+3x-x^2)/2$$

In other words, $y$ is a function of $x$. This is not a coincidence, because the ternary relation in our example had functional dependency

$$x \rightarrow y$$

If we apply Lagrange interpolation to any set of points, such as

$$(x,y) \in \{(1,1),(2,1),(3,2)\}$$

in our case, then we'll find explicit expression of $y$ as a [polynomial] function of $x$.

For centuries a function has been considered as a set of rules which describes a procedure how to transform an input to an output. Mathematicians simply refused to believe in (or saw no purpose for) functions which can't be described via nice analytic formulas. A function defined formally as a relation (set of ordered pairs) is comparatively recent (20th century) development. Having functional dependency in analytic form illuminates some classic results from database dependency theory, such as Heath's theorem.

*Heath's theorem*. Given relation $Q(x, y, z)$, and functional dependency $x \rightarrow y$, then Q can be decomposed into join of projections:

$$Q = \pi_{xy} Q \wedge \pi_{xz} Q$$

What is Heath's theorem in algebraic geometry terms?

First, let's focus on join of two relations. If both relations had the same set attributes, then the join were set intersection of tuples. A relation corresponds to a finite variety, and we already know how to perform intersection of varieties by just combining both sets of constraining equations. If two varieties have different sets of attributes as in our example, then we can expand them into larger space spanning the common set of attributes. This procedure doesn't affect the set of defining equations. For example, if we consider a variety defined by a single equation

$$(x-3)(x-2)(x-1) = 0$$

$R^1$, then it is also a variety in $R^2$ -- there is simply no constraints onto the other variable. Likewise, the variety defined by the system

$$(x-3)(x-2)(x-1) = 0$$
$$-4 + 3x - x^2 + 2y = 0$$

is actually two varieties (at least): one defined in space $R^2$ of variables $\{x, y\}$, and the other in space $R^3$ of variables $\{x, y, z\}$. Likewise, the system

$$(x-3)(x-2)(x-1) = 0$$
$$(x-3)(z-1) = 0$$
$$(z-2)(z-1) = 0$$

defines one variety in $\{x, z\}$, and the other in $\{x, y, z\}$.

As far as our example is concerned, we have achieved our goal – splitting the system of constraints into the two parts. The first system contains univariate constraint for the allowed domain $x$ values plus functional dependency. The second system contains all the equations, but the functional dependency. It is just a coincidence that none of these equations have any monomials with powers of $y$. However, if there were such monomials, we would just eliminate $y$ via substitution, leveraging the explicit formula for functional dependency.

2. IDEALS

In this section we shift the focus from varieties to dual algebraic objects — [polynomial] *ideals*. Unlike varieties ideals describe not only the set of roots where polynomial system vanishes, but also root multiplicities. In database terminology ideals are multi-relations. For example, the ideal

$$\langle x^2 - 4x + 4 \rangle$$

in the ring $k[x]$ of univariate polynomials describes unary multi-relation:

| x |
|---|
| 2 |
| 2 |

It is instructive to compare [database] relation definition with an ideal. Database relation is a set-theoretic construction involving two sets: set of attributes (relation header) and set of tuples. Then, tuples are elaborate constructs themselves (that is functions from domain to attributes). The definition for an ideal is much more concise: it is a set of multivariate polynomials which is closed with respect to addition and multiplication.

Hilbert Basis Theorem asserts that every polynomial ideal has finite basis, which is not obvious proposition given that ideals are infinite sets. It legitimizes angle bracket notation which lists basis polynomials separated with comma.

Our main focus in this section is algebra of ideals. With database theory application in mind we allow ideals from different rings. Therefore, when describing operands and result we have to be careful which polynomial ring each ideal lives in.

Consider an ideal $I$ in polynomial ring $k[x, y]$. Next, consider an ideal $J$ in polynomial ring $k[y, z]$. *The sum of ideals* $I + J$ is defined as an ideal in polynomial ring $k[x, y, z]$.

Formally,

$$I + J = \{f k_1 + g k_2 : f \in I, g \in J, k_1 \in k[x, y, z], k_2 \in k[x, y, z]\}$$

This is generalization of the standard definition of the sum of ideals living in the same ring. The proof that the result is in fact an ideal is almost verbatim. An important technicality is that we have amended standard definition of sum with factors $k_1$ and $k_2$, which helps for verification that the sum is closed under multiplication of any element of the ring $k[x,y,z]$.

The basis of the sum of ideals $I+J$ is just the concatenation of the basis of $I$ with the basis of $J$. Equivalently, the sum of an ideals $I+J$ is the smallest ideal which contains the set theoretic union $I \cup J$ of ideals (which itself is not an ideal).

The second operation we are interested in is the *intersection of ideals*, which is pure set theoretic operation. Once again, consider an ideal $I \in k[x,y]$ and an ideal $J \in k[y,z]$. Then, $I \cap J \in k[y]$. A proof that $I \cap J$ is in fact an ideal is immediate for inclusion of $0$ polynomial and closure under addition of polynomials. The proof of closure under multiplication by any element of the ring $k[y]$ follows from the fact that $I$ is closed under multiplication by $k_1 \in k[x,y]$ and, therefore, it is closed by multiplication by $k'_1 \in k[y]$, likewise, for $J$.

Since ideals are infinite sets, this definition is not practical. An alternative definition involves auxiliary variable $t$, so that given the bases of the operands $I$ and $J$ we can compute the basis of the intersection via the formula:

$$I \cap J = (tI + (1-t)J) \cap k[y]$$

Intersection of the ideal $tI + (1-t)J$ living in a larger ring $k[x,y,z,t]$ with smaller ring $k[y]$ is called *elimination ideal* (w.r.t. variables $t, x, z$).

*Proposition*. Sum and intersection of ideals are lattice operations.

This follows from the fact that we have set-theoretic intersection together with sum defined as a closure of set-theoretic union.

Axioms of relational lattice [3] provide foundation for Relational Algebra. Likewise, lattice of ideals provides a coherent foundation for algebra of multi-relations. The adjective "coherent" is the key here, because, naïve SQL implementation has always been criticized for lack of rigor and elementary inconsistencies. For example, self-join of

| x |
|---|
| 2 |
| 2 |

in MySQL would output a multi-relation containing 4 tuples, while we have just established why this operation has to be idempotent. A less sophisticated argument why self-join should have only 2 tuples involves tiny perturbation of the input:

| x |
|---|
| 1.99999 |
| 2.00001 |

From this section perspective it is evident why naïve database implementation of multi-relations via duplicated tuples is problematic – it captures significantly less information compared to a basis of an ideal.

## 3. RELATIONAL ALGEBRA WITH CAS

In this section we demonstrate that off-the-shelf *Computer Algebra System,* such as CoCoA [5], is actually a [rudimentary] RDBMS. The starting point is being able to exhibit a system of polynomial equations constraining a relation defined as a set of set of tuples. Given a binary relation with attributes $x$ and $y$ :

| x | y |
|---|---|
| 1 | 1 |
| 2 | 1 |
| 3 | 2 |

we execute the following series of CoCoA commands. First, we need to specify the polynomial ring:

```
Use XY ::= QQ[x,y];
```

Then, list the tuples:

```
Points := mat([[1, 1], [2, 1], [3, 2]]);
```

Finally, specify the ideal, and print it out:

```
I := IdealOfPoints(XY, Points);
I;
```

which outputs:

```
ideal(y^2 -3*y +2, x*y -x -3*y +3, x^2 -3*x -2*y +4)
```

Next, for the second relation over attributes $x$ and $z$ :

| X | Z |
|---|---|
| 1 | 1 |
| 2 | 1 |
| 3 | 1 |
| 3 | 2 |

we execute the series of commands

```
Use XZ ::= QQ[x,z];
Points := mat([[1, 1], [2, 1], [3, 1],[3, 2]]);
J := IdealOfPoints(XZ, Points);
J;
```

which produces

```
ideal(z^2 -3*z +2, x*z -x -3*z +3, x^3 -6*x^2 +11*x -6)
```

What is the natural join of the two relations? It is the sum $I+J$. However, we must switch to polynomial ring which contains both $\{x,y\}$ and $\{x,z\}$ :

```
Use XYZ ::= QQ[x,y,z];
I:=ideal(y^2 -3*y +2, x*y -x -3*y +3, x^2 -3*x -2*y +4);
J:=ideal(z^2 -3*z +2, x*z -x -3*z +3, x^3 -6*x^2 +11*x -6);
```
After redefining verbatim both ideals in the larger ring, we calculate their "join":

```
I+J;
```

which outputs

```
ideal(y^2 -3*y +2, x*y -x -3*y +3, x^2 -3*x -2*y +4,
z^2 -3*z +2, x*z -x -3*z +3, x^3 -6*x^2 +11*x -6)
```

This is, again an ideal of points, which is evident with the command

```
 RationalSolve(GBasis(I+J));
```

```
[[1, 1, 1], [2, 1, 1], [3, 2, 1], [3, 2, 2]]
```

Here we leveraged `GBasis` function as a way to convert an ideal into list (because `RationalSolve` accepts a list, not an ideal).

Next, projection is an *elimination ideal*. Once again, it is zero-dimensional ideal of points, so that a typical database user would like to list the tuples:

```
 RationalSolve(GBasis(elim(z, I+J)));
```

```
[[1, 1], [2, 1], [3, 2]]
```

Finally, let's hint what are the counterparts for the rest of RA operations. The set union is the intersection of an ideals. The set difference is colon ideal. The only operation which doesn't have an obvious analog in polynomial algebra is the least challenging one — renaming.

4. QUANTUM THEORY

From physicist's perspective relational databases provide classic description of the world. Surprisingly, Samson Abramsky claimed that the crux of quantum behavior – ERP paradox and Bell inequalities – can be interpreted in database terms [4]. Mathematical formulation of quantum mechanics created by Heisenberg, Born, and Jordan in 1925 involves matrices, vectors, eigenvalues, and probabilities. There can't possibly be a language more distant from relations, attributes, and values studied in the field of databases. In this section we'll investigate what does it take to consolidate frameworks of quantum and database theory. From now on, let's fix the field $k$ to be the set of complex numbers $\mathbb{C}$.

First, we shift our focus from [polynomial] ideals to their residue classes. Formally, given polynomial $p \in \mathbb{C}[x, y, z]$ its *residue class* is defined as the set

$$[p]_I = \{r \in \mathbb{C}[x, y, z] : r - p \in I\}$$

of all remainders modulo the ideal $I \subset \mathbb{C}[x, y, z]$.

Clearly, the set $[p]_I$ is obtained by adding each element of $I$ to $p$:

$$[p]_I = p + I$$

Next, the set of all residue classes is organized into a vector space. The linear operations over residue classes are defined as follows:

$$[p]_I + [r]_I = [p+r]_I$$
$$\alpha [p]_I = [\alpha p]_I$$

This implies that we must be able to choose a basis $\{[e_0]_I, [e_1]_I, ...\}$ for that vector space so that any element (residue class) is represented as

$$[p]_I = \sum_k \pi_k [e_k]_I$$

Finite varieties correspond to zero-dimensional radical ideals. Vector space of residue classes for zero-dimensional ideals is finite [2].

Now that we have inched towards quantum description and have vector space, let's discover linear operators acting on it. The critical observation is that this vector space is actually a [quotient] ring with multiplication of its element defined via

$$[p]_I [r]_I = [pr]_I$$

This operation is commutative and distributive, which helps if we want to express it in terms of basis:

$$[p]_I [r]_I = \sum_k \sum_l \pi_k [e_k]_I \rho_l [e_l]_I = \sum_k \sum_l \pi_k \rho_l [e_k e_l]_I \qquad (4.1)$$

Then, product of basis vectors themselves must be representable in terms of basis:

$$[e_k]_I [e_l]_I = [e_k e_l]_I = \sum_m \epsilon_{klm} [e_m]_I \qquad (4.2)$$

4.1 and 4.2 imply that multiplication by an element of quotient ring can be expressed as linear operator acting on the linear space of residue classes. In terms of our chosen basis $\{e_1, e_2, .., e_n\}$ it is multiplication of row vector by a matrix:

$$[p]_I \begin{pmatrix} \rho_1 \\ \rho_2 \\ \vdots \\ \rho_n \end{pmatrix} = \begin{pmatrix} p_{11} & p_{12} & \cdots & p_{1n} \\ p_{21} & & & p_{2n} \\ \vdots & & & \vdots \\ p_{n1} & p_{n2} & \cdots & p_{nn} \end{pmatrix} \begin{pmatrix} \rho_1 \\ \rho_2 \\ \vdots \\ \rho_n \end{pmatrix}$$

where each matrix element is defined as:

$$p_{ij} = \sum_k \pi_k \epsilon_{kij}$$

Let's work out multiplication matrices for the example of ideal from section 1:

$$I = \{-6 + 11x - 6x^2 + x^3, -4 + 3x - x^2 + 2y, 3 - x - 3z + xz, 2 - 3z + z^2\}$$

The vector space is 4-dimensional, so we need to choose 4 basis vectors. Let's evaluate $\{[1], [x], [x^2], [x^3]\}$ as a suitable basis. However,

$$-6[1] + 11[x] - 6[x^2] + [x^3] = [0]$$

so, the chosen vectors are linearly dependent. Either $\{[1], [x], [x^2], [z]\}$ or $\{[1], [x], [y], [z]\}$ or $\{[1], [x], [xy], [xz]\}$ is legitimate choice. The difference between the alternatives influences the amount of work required to calculate multiplication matrices. Let's fix $\{[1], [x], [y], [z]\}$ as a basis and calculate multiplication matrix for $[x]$. Multiplying by each basis vector and reducing the higher power monomials via ideal elements we get

$$[x][1] = [x] = 0[1] + 1[x] + 0[y] + 0[z]$$
$$[x][x] = [x^2] = -4[1] + 3[x] + 2[y] + 0[z]$$
$$[x][y] = [xy] = -3[1] + [x] + 3[y] + 0[z]$$
$$[x][z] = [xz] = -3[1] + [x] + 0[y] + 3[z]$$

which implies the multiplication matrix:

$$A_{[x]} = \begin{pmatrix} 0 & 1 & 0 & 0 \\ -4 & 3 & 2 & 0 \\ -3 & 1 & 3 & 0 \\ -3 & 1 & 0 & 3 \end{pmatrix}$$

Likewise, multiplication matrices for $[y]$ and $[z]$ are:

$$A_{[y]} = \begin{pmatrix} 0 & 0 & 1 & 0 \\ -3 & 1 & 3 & 0 \\ -2 & 0 & 3 & 0 \\ -2 & 0 & 1 & 2 \end{pmatrix}$$

$$A_{[z]} = \begin{pmatrix} 0 & 0 & 0 & 1 \\ -3 & 1 & 0 & 3 \\ -2 & 0 & 1 & 2 \\ -2 & 0 & 0 & 3 \end{pmatrix}$$

The *Central Theorem* of polynomial system solving by Stetter [2] asserts the following facts for 0-dimensional ideal:

- a family of multiplication matrices for ring variables is *commuting*

- this family of matrices have *joint eigenvectors*

- the ideal polynomials vanish on associated *eigenvalues*

- each of multiplication matrices can be factorized as

$$A = E \Lambda E^{-1}$$

where $E$ is matrix constructed from eigenvectors, and $\Lambda$ is diagonal matrix with eigenvalues at the main diagonal. In other words, $E$ defines a change of basis which transforms all the multiplication matrices into diagonal form.

In our example, eigenvalue problem for $A_{[x]}$ admits the following solution:

$$v_1 = \begin{pmatrix} 0 \\ 0 \\ 0 \\ 1 \end{pmatrix} \quad \lambda_1 = 3, \quad v_2 = \begin{pmatrix} 1 \\ 3 \\ 2 \\ 0 \end{pmatrix} \quad \lambda_2 = 3, \quad v_3 = \begin{pmatrix} 1 \\ 2 \\ 1 \\ 1 \end{pmatrix} \quad \lambda_3 = 2, \quad v_4 = \begin{pmatrix} 1 \\ 1 \\ 1 \\ 1 \end{pmatrix} \quad \lambda_4 = 1$$

Please note, that geometric multiplicity of eigenvalue $\lambda = 3$ is $2$, therefore, we have two-dimensional space of eigenvectors spanning $\{v_1, v_2\}$, not just one-dimensional space spanning $v_1$ and one-dimensional space spanning $v_2$. Thus, we can't construct transformation matrix $E$ yet. Since we have joint eigenvector problem, we proceed calculating eigenvectors and eigenvalues for $A_{[y]}$ :

$$v_1 = \begin{pmatrix} 0 \\ 0 \\ 0 \\ 1 \end{pmatrix} \lambda_1 = 2, \quad v_2 = \begin{pmatrix} 1 \\ 3 \\ 2 \\ 0 \end{pmatrix} \lambda_2 = 2, \quad v_3 = \begin{pmatrix} 1 \\ 0 \\ 1 \\ 1 \end{pmatrix} \lambda_3 = 1, \quad v_4 = \begin{pmatrix} 0 \\ 1 \\ 0 \\ 0 \end{pmatrix} \lambda_4 = 1$$

Finally, the eigensystem for $A_{[z]}$ :

$$v_1 = \begin{pmatrix} 1 \\ 3 \\ 2 \\ 2 \end{pmatrix} \lambda_1 = 2, \quad v_2 = \begin{pmatrix} 1 \\ 0 \\ 0 \\ 1 \end{pmatrix} \lambda_2 = 1, \quad v_3 = \begin{pmatrix} 0 \\ 0 \\ 1 \\ 0 \end{pmatrix} \lambda_3 = 1, \quad v_4 = \begin{pmatrix} 0 \\ 1 \\ 0 \\ 0 \end{pmatrix} \lambda_4 = 1$$

These individual eigenproblems can be consolidated into joint eigensolution:

$$v_1 = \begin{pmatrix} 1 \\ 3 \\ 2 \\ 2 \end{pmatrix} \lambda_1 = (3,2,2), \quad v_2 = \begin{pmatrix} 1 \\ 3 \\ 2 \\ 1 \end{pmatrix} \lambda_2 = (3,2,1), \quad v_3 = \begin{pmatrix} 1 \\ 2 \\ 1 \\ 1 \end{pmatrix} \lambda_3 = (2,1,1), \quad v_4 = \begin{pmatrix} 1 \\ 1 \\ 1 \\ 1 \end{pmatrix} \lambda_4 = (1,1,1)$$

Aggregating these eigenvectors together into the *change of basis* matrix

$$E = \begin{pmatrix} 1 & 1 & 1 & 1 \\ 3 & 3 & 2 & 1 \\ 2 & 2 & 1 & 1 \\ 2 & 1 & 1 & 1 \end{pmatrix}$$

we obtain eigendecomposition of all three multiplication matrices

$$A_{[x]} = E \begin{pmatrix} 3 & 0 & 0 & 0 \\ 0 & 3 & 0 & 0 \\ 0 & 0 & 2 & 0 \\ 0 & 0 & 0 & 1 \end{pmatrix} E^{-1}$$

$$A_{[y]} = E \begin{pmatrix} 2 & 0 & 0 & 0 \\ 0 & 2 & 0 & 0 \\ 0 & 0 & 1 & 0 \\ 0 & 0 & 0 & 1 \end{pmatrix} E^{-1}$$

$$A_{[z]} = E \begin{pmatrix} 2 & 0 & 0 & 0 \\ 0 & 1 & 0 & 0 \\ 0 & 0 & 1 & 0 \\ 0 & 0 & 0 & 1 \end{pmatrix} E^{-1}$$

In other words, we have recovered the columns of the original relation as multiplication matrices. On an afterthought, we could have just leveraged the central theorem and claimed that multiplication matrices are diagonal in *some* basis, thus avoiding tedious calculation.

In quantum mechanics physical quantities are *observables* and are formally described as linear operators. The measured values an observable are eigenvalues of corresponding linear operator. Commuting operators are observables which can be measured simultaneously.

The following *quantum-relational dictionary* summarizes this section:

| Quantum | Relational |
|---|---|
| Observable | Attribute |
| Commuting set of observables | Relation |
| Value of physical quantity (eigenvalue) | Attribute value |
| State (eigenvector) | Tuple Id (i.e. row_id) |

Foundation of quantum theory has spawned numerous research topics. One notable development was Quantum Logic originated by Birkhoff and von Neumann in the 1930s. Quantum logic is decisively different from Propositional Calculus, and its corresponding algebra is *ortholattice*. Elements of ortholattice are linear subspaces, therefore, in the context of this section, ortholattice is a structure of quotient space [of ideal residues]. Relational lattice is a structure of dual space [of ideals].